\documentstyle[pra,aps]{revtex}
\begin{document}
\draft
\title{Rotating vortex lattice in a Bose-Einstein condensate trapped in
combined  quadratic and  quartic radial  potentials}
\author{Alexander L. Fetter}
\address{Geballe Laboratory for Advanced Materials and Departments of
Physics and Applied Physics Stanford University, Stanford, CA 94305-4045}
\date{\today}
\maketitle
\begin{abstract}
A dense vortex lattice in a rotating dilute Bose-Einstein condensate is
studied with the Thomas-Fermi approximation.  The  upper critical angular
velocity $\Omega_{c2}$ occurs when the intervortex separation $b$ becomes
comparable with the vortex core radius $\xi$.  For a radial harmonic trap,
the  loss of confinement as $\Omega\to \omega_\perp$ implies a singular
behavior.  In contrast, an additional  radial quartic potential
provides a simple model for  which $\Omega_{c2}$ is readily determined.
Unlike the case of a type-II superconductor at fixed temperature, the
onset of $\Omega_{c2}$ arises  not only from  decreasing $b$ but also from
increasing
$\xi$ caused by the vanishing of the chemical potential as $\Omega\to
\Omega_{c2}$.

\end{abstract}

\section{Introduction}
The recent observations~\cite{Madi00a,Madi00b,Chev01,Abo01} of
vortex arrays and vortex lattices in  rotating dilute trapped
Bose-Einstein condensates (BEC) raise the question of analogies with
type-II superconductors.  What limits   the  number of vortices? Is there
an upper critical rotation speed
$\Omega_{c2}$ similar to the upper critical field $H_{c2}$~\cite{Tink96}
familiar from  type-II superconductors?  In the superconducting case,
the field $H_{c2}$  occurs when the distance $b\sim\sqrt{\Phi_0/\pi B}$
between vortices becomes comparable with the vortex core size $\xi\sim$
  10-100 nm in typical conventional superconducting alloys.  The measured
flux quantum
$\Phi_0=h/2e$ then implies
$H_{c2}\sim\Phi_0/\pi \xi^2\sim$ 0.1-10 T (in this limit, the distinction
between $H$ and $B$ becomes negligible).  For  rotating superfluid
$^4$He where $\xi\sim $ a few $\AA$ and
$b\sim\sqrt{\hbar/M_4\Omega}$~\cite{Donn91}, essentially the same
condition yields the  unattainably large value
$\Omega_{c2}\sim\hbar/M_4\xi^2\sim 10^{15}$ rad/s.

A similar criterion also applies to
rotating Bose condensates.  In this case,   the vortex core radius
$\xi=\hbar/\sqrt{2M\mu}\sim 0.1
\ \mu$m is macroscopic, where $\mu$ is the chemical potential,  suggesting
that   experimental study of
$\Omega_{c2}$ might well be possible.  As seen below, the situation in
rotating dilute condensates is even more favorable, because the chemical
potential decreases at large rotation speeds, increasing the vortex-core
size and thus reducing $\Omega_{c2}$.

For a trap with a radial harmonic potential $V_\perp(r_\perp) =
\frac{1}{2}M\omega_\perp^2r_\perp^2$, the behavior becomes singular when
$\Omega\to\omega_\perp$ because the outward centrifugal force
counteracts the inward force from the harmonic
trap~\cite{Butt99,Stri99,Ho01}.
Specifically, the effective  radial trap frequency
$ \left(\omega_\perp^2-\Omega^2\right)^{1/2}$
  produces an observable  centrifugal distortion of the
condensate~\cite{Rama01,Halj01,Fede01} for currently attainable values of
$\Omega/\omega_\perp\lesssim 1$.  As seen below, this behavior means that
$\omega_\perp$ effectively acts like
$\Omega_{c2}$ for pure harmonic radial confinement, and  direct
experimental  study of the limiting behavior for
$\Omega\to\omega_\perp$ would be difficult.  Thus, it is convenient to
consider an   additional stiffer radial potential, which eliminates the
singularity when $\Omega=\omega_\perp$.   In this case,
$\Omega_{c2}$ exceeds $\omega_\perp$, and the limit
$\Omega\to\Omega_{c2}$ is relatively smooth.  The  analysis is
especially simple for a  quartic radial  potential
$V_4(r_\perp) =
\frac{1}{4}kr_\perp^4$, and this example is analyzed in detail.

Section II introduces the basic Gross-Pitaevskii free energy and
condensate wave function $\Psi = |\Psi|e^{iS}$  for a rotating vortex
lattice in a trapped BEC.   In equilibrium, the vortex lattice must
experience  self-consistent solid-body rotation, which  means that
neither the phase $S$ nor the  superfluid velocity ${\bf
v}_s=(\hbar/M){\bbox\nabla} S$ can be spatially periodic in the
laboratory frame.   The transformation to the  frame rotating with
angular velocity $\bbox \Omega$ yields a   solid-body
velocity
${\bf v}_{\rm sb}={\bbox
\Omega}\times {\bf r}$  that cancels the overall rotation.  Thus, the
resulting  relative velocity ${\bf
v}_s-{\bf v}_{\rm sb}$ is indeed spatially periodic for an unbounded
vortex lattice.  This feature  facilitates a simple description of the
vortex lattice in a large trapped BEC.  The case of a trap with harmonic
radial confinement is studied in Sec.~III, and the more interesting case
of an additional  quartic radial confining potential is studied in
Sec.~IV.

\section{General formalism}

It is convenient to start from the Gross-Pitaevskii
(GP)~\cite{Gros61,Pita61} free energy in the rotating frame

\begin{equation}\label{GPen}
F=\int dV\,\left[\frac{\hbar^2}{2M}|{\bbox
\nabla}\Psi|^2+V_\perp(r_\perp)|\Psi|^2
+\case{1}{2}M\omega_z^2z^2|\Psi|^2
+\case{1}{2}g|\Psi|^4-\Psi^*{\bbox\Omega}\cdot {\bf r\times
p}\,\Psi\right],
\end{equation}
where $V_\perp(r_\perp)$ is the transverse radial confining potential
and $g=4\pi a\hbar^2/M$ relates the interparticle coupling constant to
the $s$-wave scattering length $a$ (here assumed positive).
The  representation
$\Psi = |\Psi| e^{iS}$ emphasizes the hydrodynamic aspects of the
behavior, with condensate density $n_0=|\Psi|^2$
and superfluid velocity
${\bf v}_s=\hbar{\bbox \nabla}S/M$. The quantity $-\Psi^*\,{\bbox
\Omega}\cdot {\bf r\times p}\Psi$ can be written  as
$\case{1}{2}i\hbar \Omega\,\partial\, |\Psi|^2/\partial \phi-M\,{\bf
v}_{\rm  sb}\cdot{\bf v}_s \,|\Psi|^2$, and  the  first term makes no
contribution  to  the spatial integral.  Straightforward manipulations of
Eq.~(\ref{GPen}) yield

\begin{equation}\label{GPen1}
F=\int dV\,\left[\case{1}{2}M\left({\bf v}_s-{\bf
v}_{\rm sb}\right)^2|\Psi|^2 +
\frac{\hbar^2}{2M}\left({\bbox
\nabla}|\Psi|\right)^2 +V_\perp(r_\perp)|\Psi|^2+V_{\rm
cent}(r_\perp ) |\Psi|^2
+\case{1}{2}M\omega_z^2z^2|\Psi|^2 +\case{1}{2}g|\Psi|^4\right],
\end{equation}
where $V_{\rm cent}(r_\perp)= -\frac{1}{2}M\Omega^2 r_\perp^2$ is an
effective repulsive centrifugal potential.

The simplest variational
model is to take the phase
$S$ as that for a classical  unbounded vortex lattice (with vortices
aligned along
$\hat z$), which has been analyzed in detail by Tkachenko~\cite{Tkac65}
(as a model for a vortex lattice in rotating superfluid $^4$He).   In
this case, the total superfluid velocity  includes the
divergent  locally axisymmetric circulating flow  near each vortex core.
The resulting singular kinetic energy in Eq.~(\ref{GPen1}) is cut
off by a self-consistent core structure of characteristic dimension $\xi
=
\hbar/\sqrt{2M\mu}$, where $\mu $ is the chemical
potential~\cite{Gros61,Pita61,Fett01}.

  This classical hydrodynamic phase
$S$ cannot be a spatially  periodic function in the $xy$ plane
because the vortex array  induces a self-consistent  rotation.
Instead, this phase  obeys definite
quasiperiodic conditions~\cite{Tkac65,Fett83} that depend on the details
of the lattice structure.  For the same reasons,
${\bf v}_s
\propto {\bbox\nabla}S$ is also not spatially periodic.  In contrast,  the
quantity
${\bf v}_s-{\bf v}_{\rm sb}$ (namely the superfluid velocity as seen in
the rotating frame) {\em is} spatially periodic, because the subtracted term
${\bf v}_{\rm sb}$ cancels  the effect of the quasiperiodic terms in
${\bbox\nabla} S$,  considerably simplifying  the subsequent analysis.

According to Feynman's picture of a rotating superfluid~\cite{Feyn55},
the vortices have a uniform areal  density $n_v = 2M\Omega/h$,
  ensuring that the mean vorticity of the vortex lattice mimics that of
solid-body rotation
${\bbox
\nabla}\times {\bf v}_{\rm sb}= 2{\bbox\Omega}$.  Thus the area per
vortex $ n_v^{-1} \equiv \pi b^2$ can be taken to define an intervortex
separation $b = \sqrt{\hbar/M\Omega}$;  this characteristic length sets
the scale of the vortex lattice.  For the present purposes, the detailed
lattice structure is unimportant (for example, the  free
energies of the triangular and square configurations have the same
logarithmic contributions $\propto \ln(b/\xi)$ and differ only in the
  additive constants~\cite{Tkac66}); it will be convenient  to
use a Wigner-Seitz approximation in which each polygonal unit cell is
replaced by an equivalent circular cell of radius $b$.  Evidently, the
ratio $\xi^2/b^2$ characterizes the fractional volume occupied by the
``normal'' vortex cores.

For any reasonable transverse confining potential $V_\perp(r_\perp)$, the
single-particle ground state will have some characteristic transverse
dimension
$d_\perp$, along with the characteristic axial dimension $d_z
=\sqrt{\hbar/M\omega_z}$ that is set by the axial harmonic confining
potential in Eqs.~(\ref{GPen}) and (\ref{GPen1}).     Let $N_0$ be the
number of atoms in the condensate at low temperature.  In the Thomas-Fermi
(TF) limit~\cite{Baym96}, the condensate density is taken as locally
constant, which holds when the dimensionless interaction parameter
$N_0a/d_\perp$ is large.  In this case, the repulsive Hartree (mean-field)
interactions  expand the condensate relative to its noninteracting
transverse and axial   dimensions $d_\perp$ and
$d_z$. In particular, the radial and axial condensate radii $R_\perp $
and $R_z$ are simply the classical turning points for a particle of
energy $\mu$.  Thus, $R_z=\sqrt{2\mu/M\omega_z^2}$ depends only on
$\mu$, but
$R_\perp$ also depends on $\Omega$ because of the repulsive centrifugal
potential
$V_{\rm cent}(r_\perp)$
  in Eq.~(\ref{GPen1}).

  For a rotating condensate with
chemical potential
$\mu$, the GP equation derived from  Eq.~(\ref{GPen1}) implies the
corresponding TF density

\begin{equation}\label{dens}
  |\Psi(r_\perp,z)|^2 = \frac{\mu}{g}\left[1 -
\frac{V_\perp(r_\perp)}{\mu}+
\frac{\Omega^2r_\perp^2}{2\mu}-\frac{M\omega_z^2z^2}{2\mu} -
\frac{M\left({\bf v}_s-{\bf
v}_{\rm sb}\right)^2}{2\mu}\right],
\end{equation}
obtained by omitting the gradient of the condensate density.  If the
vortex cores do not overlap significantly (so that $\xi^2/b^2\ll 1$),
the condensate density can be approximated by a product
$|\Psi(r_\perp,z)|^2 \approx |\Psi_{TF}(r_\perp,z)|^2\, u^2(r_\perp)$ of
the TF density

\begin{equation}\label{TFdens}
  |\Psi_{TF}(r_\perp,z)|^2 = \frac{\mu}{g}\left[1 -
\frac{V_\perp(r_\perp)}{\mu}+
\frac{\Omega^2r_\perp^2}{2\mu}-\frac{M\omega_z^2z^2}{2\mu} \right]
\end{equation}
  in the absence of the vortex lattice, and a factor

\begin{equation}\label{period}
u^2(r_\perp)= 1-
   \frac{M\left({\bf v}_s-{\bf
v}_{\rm sb}\right)^2}{2\mu}
\end{equation}
that is spatially periodic in the transverse plane and must be cut off
near  the vortex core to ensure that $u^2$ remains
positive.  In any given unit cell, it
has the local form

\begin{equation}\label{period1}
u^2(r_\perp)= 1-
\frac{\xi^2}{r_\perp^2}+\frac{2\xi^2}{b^2}-\frac{\xi^2r_\perp^2}{b^4},
\end{equation}
where a  circular Wigner-Seitz cell has been used with $\xi\le r_\perp\le
b$. In effect, the resulting density  is that of a vortex-free condensate
with narrow holes  along the axes of the vortex
lattice~\cite{Abo01,Fede01,Voss01}.

The chemical potential is determined by the general  condition  that
$N_0 = \int dV \,|\Psi(r_\perp,z)|^2$.  Here the factorized form yields

\begin{eqnarray}
N_0 &\approx& \int dV\, |\Psi_{TF}(r_\perp,z)|^2\, u^2(r_\perp)\nonumber
\\
&=&\frac{\mu M R_z}{3\pi a\hbar^2}\int
d^2r_\perp\,\left[1+\frac{M\Omega^2r_\perp^2}{2\mu}
-\frac{V_\perp(r_\perp)}{\mu}
\right]^{\!3/2}\,u^2(r_\perp).\label{TFsum}
\end{eqnarray}
    For a dense vortex lattice,  the intervortex separation
$b$ is small compared to the   TF transverse radius  $R_\perp$, so that
the first  factor varies  slowly on the length scale $b$.  Thus, $u^2$ can
be averaged   over any single unit cell, and the Wigner-Seitz
approximation~\cite{Fett83} in Eq.~(\ref{period1}) yields

\begin{equation}
\langle u^2\rangle =\frac{1}{\pi b^2} \int_{cell} d^2
r_\perp\,u^2(r_\perp)\approx
1-2\frac{\xi^2}{b^2}\left[\ln\left(\frac{b}{\xi}\right)
-\case{3}{4}\right];
\end{equation}
as anticipated, $\langle u^2\rangle \approx 1$ for $\xi\ll b$.  In the
limit of a large condensate with many  vortices,
Eq.~(\ref{TFsum}) can therefore be approximated by

\begin{equation}\label{TFint}
N_0 = \frac{\mu M R_z \langle u^2\rangle}{3\pi a\hbar^2 }\int d^2r_\perp\,
\left[1+\frac{M\Omega^2r_\perp^2}{2\mu}
-\frac{V_\perp({ r}_\perp)}{\mu}
\right]^{\!3/2}.
\end{equation}

\section{radial harmonic confining potential}

The simplest situation is the harmonic radial potential
$V_\perp(r_\perp)=
\case{1}{2}M\omega_\perp^2 r_\perp^2$, in which case Eq.~(\ref{TFint})
implies the TF  condensate radius

\begin{equation}\label{Rharm}
R_\perp = \sqrt{\frac{2\mu}{M(\omega_\perp^2-\Omega^2)}}.
\end{equation}
Familiar manipulations
yield

\begin{equation}\label{mu}
\frac{\mu}{\hbar\omega_\perp}=
\frac{1}{2}\left(\frac{15N_0a\lambda}{d_\perp\langle
u^2\rangle}\right)^{\!\!2/5}
\left(1-\frac{\Omega^2}{\omega_\perp^2}\right)^{\!\!2/5},
\end{equation}
where $\lambda = \omega_z/\omega_\perp$ is the axial anisotropy
parameter.

Except for a very narrow region $\Omega/\omega_\perp\lesssim 1$ close to
the singular limit,  the vortex cores occupy
negligible volume, so that  $\langle u^2\rangle\approx 1$.  In this case,
the condensate particle number $N_0$ remains essentially constant, and
Eq.~(\ref{mu}) shows how the chemical potential decreases with increasing
rotation speed
$\overline \Omega\equiv\Omega/\omega_\perp$.  In this same large interval,
  Eq.~(\ref{Rharm})  implies that

\begin{equation}
\frac{R_\perp^2(\Omega)}{d_\perp^2}=\left(\frac{15N_0a\lambda}{d_\perp}
\right)^{\!\!2/5}
{\left(1-\overline \Omega^2\right)^{-3/5}}.
\end{equation}
Equivalently,
\begin{equation}
\frac{R_\perp(\Omega)}{R_\perp(0)}=
{\left(1-\overline\Omega^2\right)^{-3/10}},
\end{equation}
as seen  in recent experiments~\cite{Rama01}.

  The fraction of depleted condensate occupied by the vortex cores is given
quite generally by
$\xi^2/b^2=\hbar\Omega/2\mu$, and use of  Eq.~(\ref{mu}) gives
the specific result

\begin{equation}\label{frac}
\frac{\xi^2}{b^2}=\frac{\hbar\Omega}{2\mu}=
\frac{\overline \Omega}{2}\,
\frac{\epsilon^{2/5}}
{\left(1-\overline \Omega^2\right)^{2/5}}
\end{equation}
where

\begin{equation}\label{eps}
\epsilon \equiv \frac{d_\perp\langle
u^2\rangle}{15 N_0a\lambda}\approx \frac{d_\perp^5}{R_\perp^5(0)}\ll 1
\end{equation}
is small in the TF limit.
Equation (\ref{frac}) shows that the ratio $\xi^2/b^2$ increases linearly
with $\overline \Omega$ but remains small (of order
$\epsilon^{2/5}$) until $1-\overline \Omega \lesssim
\epsilon/8\sqrt2$, when the ratio  grows rapidly toward~1. If
$\Omega_{c2}$  is defined like
$H_{c2}$ for a type-II superconductor, $\Omega_{c2}$ corresponds to the
limit
$\xi\sim b$. As a result,  a radial harmonic  trap potential leads to
  singular behavior when $\Omega\to \omega_\perp$~\cite{Ho01}.
Specifically, the disappearance of the condensate  associated with the
sudden expansion of the ``normal'' vortex cores presumably implies a
phase transition, which  occurs essentially
simultaneously with the loss of confinement as
$\overline \Omega\to
1$~\cite{Butt99,Stri99,Ho01,Fede01}.

This straightforward analysis indicates that any study of a dense vortex
lattice near $\Omega_{c2}$ for a radial harmonic trap will encounter
significant difficulties arising from the softening of the effective trap
potential.  As a result, it is interesting to consider a stiffer radial
potential, and the next section treats one possible example.

\section{radial quartic confining potential}

As a specific example, let $V_\perp (r_\perp)$ include a quartic
radial trap potential $V_4(r_\perp)  = \frac{1}{4}kr_\perp^4$ in addition
to the quadratic potential
$\case{1}{2}M\omega_\perp^2 r_\perp^2$;
Lundh~\cite{Lund01} has recently examined this and other power-law
potentials in a theoretical study of the formation of a multiply
quantized vortex in a rotating condensate.  For a two-dimensional ideal
gas confined with pure $V_4(r_\perp)$, the balance between the
ground-state kinetic energy
$\sim
\hbar^2/Md_4^2$ and ground-state potential energy $\sim kd_4^4$
implies a ground-state mean radius
$d_4\sim (\hbar^2/Mk)^{1/6}$ and ground-state energy
$E_4=\hbar\omega_4\sim (\hbar^4k/M^2)^{1/3}$.  For definiteness,
it is convenient to ignore numerical constants of order unity and take

\begin{equation}\label{quartic}
  E_4 = \hbar\omega_4 = \frac{\hbar^2}{Md_4^2} =
kd_4^4,
\end{equation}
as the relevant dimensional quantities.

For a rotating condensate in this combined radial trap, Eq.~(\ref{TFint})
becomes

\begin{equation}\label{TFint4}
N_0 = \frac{\mu M R_z \langle u^2\rangle}{3 a\hbar^2 }\int_0^{R_\perp^2}
du\,
\left[1-\frac{M(\omega_\perp^2-\Omega^2)}{2\mu}\,u
-\frac{k }{4\mu}\,u^2
\right]^{\!3/2},
\end{equation}
where $u = r_\perp^2$, and $R_\perp^2$ is the turning point where the
integrand vanishes.  The presence of the quartic potential means that the
external rotation speed $\Omega$ can  now exceed $\omega_\perp$.  In this
regime ($\overline \Omega >~1$),  the particle  density actually attains a
local minimum on  the axis of symmetry, but this effect is probably
difficult to detect.

The integral in Eq.~(\ref{TFint4}) can be expressed in terms of a
dimensionless parameter

\begin{equation}\label{eta}
\eta\equiv \frac{M\left(\omega_\perp^2-\Omega^2\right)}{2\sqrt{k\mu}}.
\end{equation}
A detailed analysis yields   the final form

\begin{equation}\label{int}
\int_0^{R_\perp^2}
du\,
\left[1-\frac{M(\omega_\perp^2-\Omega^2)}{2\mu}\,u
-\frac{k }{4\mu}\,u^2
\right]^{\!3/2} = 2\sqrt{\frac{\mu}{k}}\,f(\eta),
\end{equation}
where

\begin{equation}\label{f}
f(\eta) =
\frac{3\pi}{16}\left(1+\eta^2\right)^2\,\left[1-\frac{2}{\pi}
\arcsin\left(\frac{\eta}{\sqrt{1+\eta^2}}\right)\right]
-\frac{3\eta^3}{8}-\frac{5\eta}{8}.
\end{equation}
It has the limiting forms

\begin{equation}\label{limit}
f(\eta)\approx\cases{(5\eta)^{-1}&for $\eta\gg 1$,\cr
                     \frac{3}{16}\pi -\eta& for $|\eta|\ll 1$,\cr
                       \frac{3}{8}\pi \eta^4& for $\eta\ll -1$.\cr}
\end{equation}

Equations (\ref{TFint4}) and (\ref{int}) can be combined with
$R_z=\sqrt{2\mu/M\omega_z^2}$ to express the chemical potential in terms of the
parameter
$\eta$

\begin{equation}\label{mu4}
\left(\frac{\mu}{\hbar\omega_\perp}\right)^{\!\!
2}=\frac{N_0a}{d_\perp\langle
u^2\rangle}\,\frac{\omega_z}{\omega_\perp}\,\frac{3}{2\sqrt
2 \,f(\eta)}\,\left(\frac{d_\perp}{d_4}\right)^{\!\! 3},
\end{equation}
where $d_\perp/d_4$ is a dimensionless measure of the relative strength
of the quartic potential (note that $d_\perp/d_4\to 0$ in the limit that
the quartic coupling constant $k$ becomes small).  In addition,
Eq.~(\ref{eta}) can be rewritten to express the rotation speed as a
function of $\mu$ and $\eta$

\begin{equation}\label{Omega4}
\overline \Omega^2 \equiv \frac{\Omega^2}{\omega_\perp^2}= 1-2\eta
\,\sqrt{\frac{\mu}{\hbar\omega_\perp}}\,
\left(\frac{d_\perp}{d_4}\right)^{\!\!3};
\end{equation}
Note that $\eta$ is large and positive for a nonrotating condensate
($\Omega=0$), $\eta$  vanishes for $\overline \Omega=1$, and $\eta$
becomes negative for $\overline \Omega>1$ (this limit can occur only
for nonzero positive $d_\perp/d_4$).  These two equations provide a parametric
representation of the dependence of the dimensionless chemical potential
$\overline
\mu=\mu/\hbar\omega_\perp$ on the rotation speed, generalizing the result in
Eq.~(\ref{mu}) for a pure harmonic potential to include the effect of an
additional  quartic potential.

It is not difficult to see from Eqs.~(\ref{limit})  and (\ref{mu4}) that
$\mu$ vanishes like $\eta^{-2}$ as $\eta\to -\infty$, so that the
right-hand side of Eq.~(\ref{Omega4}) remains finite in the same limit.  Thus
the chemical potential vanishes when the  rotation frequency attains its
maximum value.  This maximum rotation speed can be identified as the upper
critical value $\Omega_{c2}$  because the
   vortex core size $\xi=\hbar/\sqrt{2M\mu}$ diverges as $\mu \to 0$.  A
detailed analysis (approximating $\langle u^2\rangle\approx 1$) shows that

\begin{equation}\label{c2}
\Omega_{c2}^2 \approx \omega_\perp^2 +\omega_4^2\left(\frac{32\sqrt
2}{\pi}\,\frac{N_0a}{d_4}\,\frac{\omega_z}{\omega_4}\right),
\end{equation}
where the positive quantity  $\Omega_{c2}^2-\omega_\perp^2$ has been expressed
solely in terms of the parameters associated with the quartic term in the
confining potential.  Figure \ref{fig1} shows the dimensionless chemical
potential $\overline\mu$ as a function of the dimensionless rotation speed
$\overline
\Omega$ for two illustrative cases (note that $\overline \mu$ remains
positive for $\overline \Omega=1$).

In the presence of the additional quartic confining potential, the TF radius
$R_\perp$ is given by

\begin{equation}
\frac{R_\perp^2}{d_\perp^2} =
2\sqrt{\frac{\mu}{\hbar\omega_\perp}}\,
\left(\frac{d_4}{d_\perp}\right)^{\!\!3}\,\frac{1}{\sqrt{1+\eta^2}+\eta}.
\end{equation}
A combination with Eq.~(\ref{Omega4}) provides a parametric relation for
$R_\perp(\Omega)$.  Figure \ref{fig2} shows the dimensionless ratio
$R_\perp^2/d_\perp^2$ as a function of the dimensionless rotation speed
$\overline \Omega$ (note that $R_\perp^2/d_\perp^2$ remains finite as
$\overline \Omega\to\overline \Omega_{c2}$).

Finally, the ratio of the vortex core radius to the intervortex separation
follows from

\begin{equation}
\frac{\xi^2}{b^2} = \overline\Omega\,\frac{\hbar\omega_\perp}{2 \mu} =
\frac{\overline\Omega}{2\overline\mu},
\end{equation}
and use of Eq.~(\ref{Omega4}) to eliminate $\eta$ in favor of
$\overline\Omega$ gives the generalization of Eq.~(\ref{frac}) for the case of
a combined quadratic and quartic confining potential.  Figure \ref{fig3}
shows the $\overline\Omega$ dependence of $\xi^2/b^2$ for two illustrative
cases;  the sharp increase  near
$\overline\Omega_{c2}$ and the finite limiting value are particularly
evident.

\section{Discussion and conclusions}

The behavior of a dense vortex lattice in a dilute trapped rotating
Bose-Einstein condensate has acquired a new interest because of recent
observations of such lattices~\cite{Madi00b,Chev01,Abo01}.  The
experiments use radial harmonic traps, and the loss of confinement as the
external rotation speed $\Omega$ approaches the trap frequency
$\omega_\perp$ implies that the behavior becomes singular~\cite{Ho01}.  As a
result, it would be difficult to study the  approach to what is effectively the
upper critical angular velocity
$\Omega_{c2}\sim \hbar/M\xi^2$ when the vortex cores overlap.  The
present analysis considers the addition of a stiffer quartic potential
$V_\perp(r_\perp)=\frac{1}{4}kr_\perp^4$, which ensures confinement for any
$\Omega$.  Thus, the approach to $\Omega_{c2}$ is more gradual and  could
well be observed.  In particular,
$\Omega_{c2}/\omega_\perp$  exceeds 1 and should be accessible to
experiments.  In principle, a similar analysis is possible for other
stiff radial confining potentials, for example, other power laws or an optical
dipole waveguide made from a hollow blue-detuned laser beam~\cite{Bong01}.

The Thomas-Fermi approximation assumes that the total kinetic energy of
the condensate is much smaller than the energies associated with the
external trapping potential and the interparticle Hartree potential.  This
picture certainly applies for a relatively small number of vortices when
$\xi\ll b\ll R_\perp$, but it fails near  $\Omega_{c2}$ when
$\xi \lesssim b$ because of the density gradient  near the many
vortex cores.  This effect will not qualitatively alter the value of
$\Omega_{c2}$ determined for the quartic potential, but it will affect the
detailed description for
$\Omega\lesssim
\Omega_{c2}$.   This interesting question remains open.

Another question concerns the number of
atoms $N_0$ in the condensate, which here has been assumed to  remain
fixed.  This picture also must fail sufficiently near
$\Omega_{c2}$ because the vortex cores fill the entire volume.  For type-II
superconductors~\cite{Tink96} at a given temperature, the vortex core size
$\xi$ remains fixed.  In that system, an increased applied magnetic field
reduces the intervortex separation $b$, leading to the disappearance of
the superconducting component because
  the ``normal'' cores eventually  overlap.  In a dilute Bose-Einstein
gas, however, the core size
$\xi = \hbar/\sqrt{2M\mu}$ itself increases and diverges as
$\Omega\to\Omega_{c2}$ because of the decreasing chemical potential.
Thus the approach to the upper critical angular velocity is  more
sudden in a dilute trapped gas.  Unfortunately, inclusion of the
quantum depletion is fairly complicated for a spatially nonuniform
medium~\cite{Sing96,Hutc97,Dalf97,Fett98,Dalf99} although the general formalism
is well known~\cite{Huge59,Fett71}.

At zero temperature in the grand canonical
ensemble for a rotating system in equilibrium at chemical potential $\mu$
and
  an angular velocity $\Omega$, for example, the ground-state
expectation value of  the operator
$\hat K =
\hat H-\mu\hat N-\Omega \hat L_z$ is the
  thermodynamic potential $\tilde F(\mu,\Omega)= \langle \hat K \rangle$.
In the presence of a Bose-Einstein condensate with $N_0$ condensate atoms,
this function also depends on $N_0$, so that $ \langle \hat K \rangle =
\tilde F(\mu,\Omega;N_0)$.  The usual thermodynamic relation
$N(\mu,\Omega;N_0) = -(\partial\tilde F/\partial \mu)_{\Omega
N_0}$ determines the mean number of particles, which here depends not only
on
$\mu $ and
$\Omega$, but also on $N_0$.  This latter parameter can be fixed by
adjusting  $N_0$ to minimize $\tilde F$, so that $(\partial \tilde
F/\partial N_0)_{\mu\Omega}=0$.  For a uniform Bose gas in a stationary box
of volume $V$, it is straightforward to verify that this procedure yields
the familiar zero-temperature depletion of the condensate $(N-N_0)/N\approx
\case{8}{3}(na^3/\pi)^{1/2}$ as well as the first
correction to the chemical potential~\cite{Huge59,Fett71,Bogo47}. It
should be feasible to extend this analysis to a dilute rotating trapped
Bose condensate, and this problem definitely merits further study.

\acknowledgments{I thank D.~Feder, M.~Linn and A.~Svidzinsky for valuable
comments and suggestions.  This work is supported in part by the  National
Science Foundation under Grant No.~DMR 99-71518.}

\begin{figure}
\caption{Dimensionless chemical potential $\overline\mu
\equiv\mu/\hbar\omega_\perp$ for the combined quadratic and quartic radial
confining potential as a function of the dimensionless  angular velocity
$\overline \Omega=\Omega/\omega_\perp$ with $(N_0
a/d_\perp)\,(\omega_z/\omega_\perp)= 10 ^4$. The two curves correspond
to the values
   (a) $d_4/d_\perp= 5$ (b) $d_4/d_\perp= 2$. Note that $\overline \mu$ remains
finite at $\overline \Omega = 1$ for both values of $d_4/d_\perp$.}
\label{fig1}
\end{figure}

\begin{figure}
\caption{Dimensionless squared radius  $R_\perp^2/d_\perp^2$
for the combined quadratic and
quartic radial confining potential as a
function of the dimensionless  angular velocity
$\overline \Omega=\Omega/\omega_\perp$ with $(N_0
a/d_\perp)\,(\omega_z/\omega_\perp)= 10 ^4$. The two curves correspond
to the values
   (a) $d_4/d_\perp= 5$ (b) $d_4/d_\perp= 2$. In contrast to the behavior
seen in Eq.~(\ref{Rharm}) for a
rotating harmonic radial potential, the squared radius here
remains finite for both values of $d_4/d_\perp$  as
$\Omega\to\Omega_{c2}$ given in Eq.~(\ref{c2}).}
\label{fig2}
\end{figure}

\begin{figure}
\caption{Fraction of volume $\xi^2/b^2$ occupied by the vortex cores for
the combined quadratic and quartic radial confining potential as a
function of the dimensionless  angular velocity
$\overline \Omega=\Omega/\omega_\perp$ with $(N_0
a/d_\perp)\,(\omega_z/\omega_\perp)= 10 ^4$. The two curves correspond
to the values
   (a) $d_4/d_\perp= 5$ (b) $d_4/d_\perp= 2$.  Note that $\xi^2/b^2$ remains
finite for both values of
$d_4/d_\perp$ as
$\Omega\to\Omega_{c2}$ given in Eq.~(\ref{c2}).}
\label{fig3}
\end{figure}


\begin{references}

\bibitem{Madi00a}  K.~W.~Madison, F.~Chevy, W.~Wohlleben, and
J.~Dalibard, Phys.~Rev.~Lett.~{\bf 84}, 806 (2000).

\bibitem{Madi00b}  K.~W.~Madison, F.~Chevy, W.~Wohlleben, and
J.~Dalibard, J.~Mod.~Opt.~{\bf 47}, 2715 (2000).

\bibitem{Chev01}  F.~Chevy, K.~W.~Madison, V.~Bretin, and J.~Dalibard,
e-print: cond-mat/0104218.

\bibitem{Abo01} J.~R.~Abo-Shaeer, C.~Raman, J.~M.~Vogels, and
W.~Ketterle, Science {\bf 292}, 476 (2001).

\bibitem{Tink96}  See, for example, M.~Tinkham, {\it Introduction to
Superconductivity\/}, second edition (McGraw-Hill, New York, 1996),
Secs.~4.8 and 4.11.

\bibitem{Donn91}  R.~J.~Donnelly, {\it Quantized Vortices in Helium
II\/} (Cambridge University Press, Cambridge, 1991), Chaps.~4 and 5.

\bibitem{Butt99} D.~A.~Butts and D.~S.~Rokhsar, Nature {\bf 397}, 327
(1999).

\bibitem{Stri99}  S.~Stringari,  Phys.~Rev.~Lett.~{\bf 82}, 4371 (1999).

\bibitem{Ho01}  T.-L.~Ho, Phys.~Rev.~Lett.~{\bf 87}, 060403 (2001)

\bibitem{Rama01} C.~Raman, J.~R.~Abo-Schaeer, J.~M.~Vogels, K.~Xu, and
W.~Ketterle, e-print: cond-mat/0106235.

\bibitem{Halj01} P.~C.~Haljan, I.~Coddington, P.~Engels, and
E.~A.~Cornell, e-print: cond-mat/0106362.

\bibitem{Fede01}  D.~L.~Feder and C.~W.~Clark, e-print: cond-mat/0108019.

\bibitem{Gros61}  E.~P.~Gross, Nuovo Cimento {\bf 20}, 454 (1961).

\bibitem{Pita61}   L.~P.~Pitaevskii, Zh.~Eksp.~Teor.~Fiz.~{\bf 40}, 464
(1961) [Sov.~Phys.--JETP {\bf 13}, 451 (1961)].

\bibitem{Tkac65}  V.~K.~Tkachenko, Zh.~Eksp.~Teor.~Fiz.~{\bf 49}, 1875
(1965) [Sov.~Phys.--JETP {\bf 22}, 1282 (1966)].

\bibitem{Fett01}  For a recent review, see A.~L.~Fetter and
A.~A.~Svidzinsky, J.~Phys.~Cond.~Mat.~{\bf 13}, R135 (2001).

\bibitem{Fett83}  A.~L.~Fetter, J.~A.~Sauls, and D.~L.~Stein,
Phys.~Rev.~B {\bf 28}, 5061 (1983).

\bibitem{Feyn55}  R.~P.~Feynman, in {\it Progress in Low Temperature
Physics}, edited by C.~J.~Gorter, Vol.~1 (North-Holland, Amsterdam,
1955), p.~17.

\bibitem{Tkac66}  V.~K.~Tkachenko, Zh.~Eksp.~Teor.~Fiz.~{\bf 50}, 1573
(1966) [Sov.~Phys.--JETP {\bf 23}, 1049 (1966)].

\bibitem{Baym96}  G.~Baym and C.~J.~Pethick, Phys.~Rev.~Lett.~{\bf 76},
6 (1996).

\bibitem{Voss01}   D.~Voss, Science {\bf 291}, 2301 (2001).

\bibitem{Lund01}  E.~Lundh, e-print: cond-mat/0103272.

\bibitem{Bong01}  K.~Bongs, S.~Burger, S.~Dettmer, D.~Hellweg, J.~Arlt,
W.~Ertmer, and K.~Sengstock, Phys.~Rev.~A {\bf 63}, 031602(R) (2001).

\bibitem{Sing96}  K.~G.~Singh and D.~S.~Rokhsar, Phys.~Rev.~Lett.~{\bf 77},
1667 (1996).

\bibitem{Hutc97}  D.~A.~W.~Hutchinson, E.~Zaremba, and A.~Griffin,
  Phys.~Rev.~Lett.~{\bf 78}, 1842 (1997).

\bibitem{Dalf97}  F.~Dalfovo, S.~Giorgini, M.~Guilleumas, L.~Pitaevskii, and
S.~Stringari, Phys.~Rev.~A {\bf 56}, 3840 (1997).

\bibitem{Fett98}  A.~L.~Fetter and D.~Rokhsar, Phys.~Rev.~A {\bf 57}, 1191
(1998).

\bibitem{Dalf99}  F.~Dalfovo, S.~Giorgini, L.~P.~Pitaevskii, and
S.~Stringari, Rev.~Mod.~Phys.~{\bf 71}, 463 (1999), Sec.~III.E.

\bibitem{Huge59}  N.~M.~Hugenholtz and D.~Pines, Phys.~Rev.~{\bf 116}, 489
(1959).

\bibitem{Fett71}  A.~L.~Fetter and J.~D.~Walecka, {\it Quantum Theory of
Many-Particle Systems} (McGraw-Hill, New York, 1971), Sec. 18.

\bibitem{Bogo47} N.~N.~Bogoliubov, J.~Phys.~(USSR) {\bf 11}, 23 (1947).





\end{references}
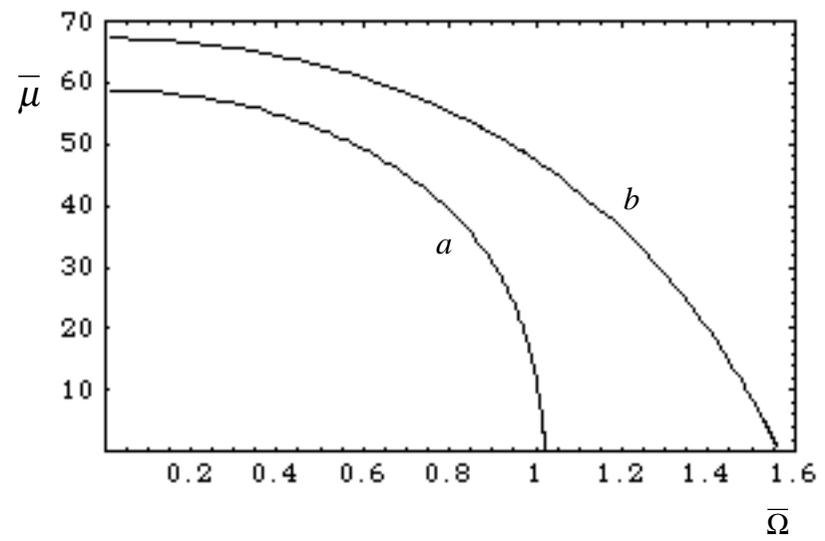
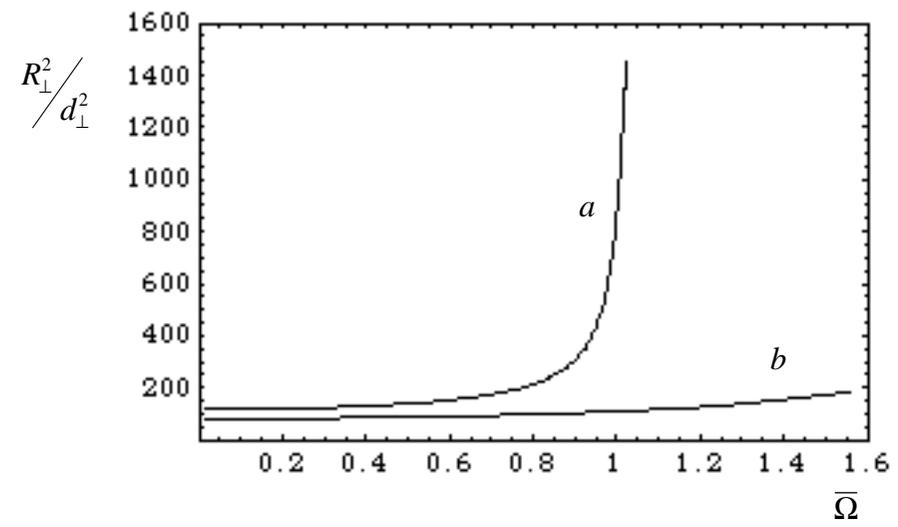
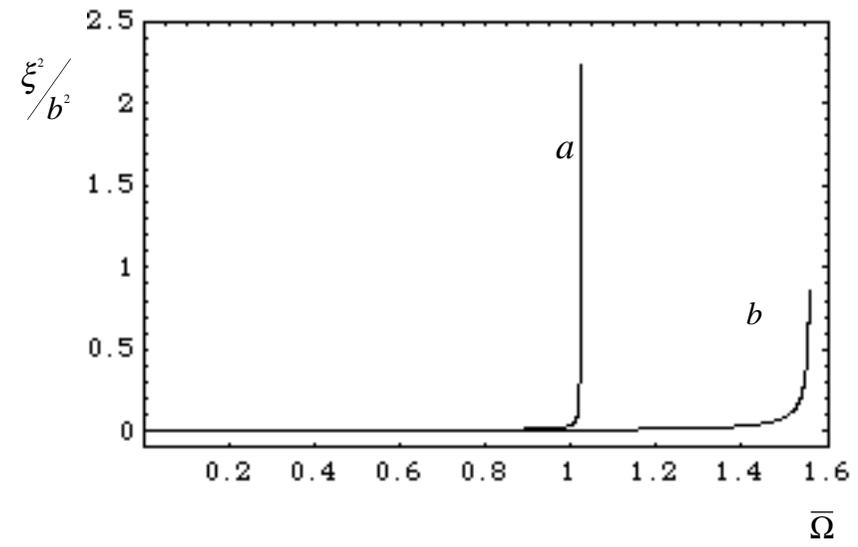
\end{document}